# Calibration of Machine Learning Classifiers for Probability of Default Modelling

Pedro G. Fonseca and Hugo D. Lopes
*James Finance (CrowdProcess Inc.)*
*October 24th, 2017*

## Abstract

Binary classification is highly used in credit scoring in the estimation of probability of default. The validation of such predictive models is based both on rank ability, and also on calibration (i.e. how accurately the probabilities output by the model map to the observed probabilities). In this study we cover the current best practices regarding calibration for binary classification, and explore how different approaches yield different results on real world credit scoring data. The limitations of evaluating credit scoring models using only rank ability metrics are explored.
A benchmark is run on 18 real world datasets, and results compared. The calibration techniques used are Platt Scaling and Isotonic Regression. Also, different machine learning models are used: Logistic Regression, Random Forest Classifiers, and Gradient Boosting Classifiers. Results show that when the dataset is treated as a time series, the use of re-calibration with Isotonic Regression is able to improve the long term calibration better than the alternative methods. Using re-calibration, the non-parametric models are able to outperform the Logistic Regression on Brier Score Loss.

# Table of contents

# 1. Introduction

## 1.1. Probability of Default modelling

In many applications it is important to correctly establish the probability of the occurrence of an event. One such application is in Credit Scoring, where lenders can use a classification system, which can range from a simple scorecard to a complex machine learning algorithm, to attribute a certain rating to each loan application (Anderson, 2007; Khandani, 2010). The rating grade attributed by the lender can then be transformed into an estimated Probability of Default (PD). The correct mapping between rating grades and PDs constitutes the PD calibration. This subject, while crucial to the accurate validation of models, is often mentioned in literature, but receives much less attention than the rank / ordering metrics (e.g. Gini / Kolmogorov Smirnov).

Machine learning models are increasingly used by lenders as part of the credit attribution process (Hue, 2017). When these machine learning models are used for calibration, they do not necessarily produce calibrated probabilities (Caruana, 2016), leading to the need for calibration. Most academic research on calibration tends to focus on clean and relatively balanced datasets, while in reality lending datasets are often highly imbalanced and with noisy data. This paper aims to benchmark the calibration of different models on retail lending datasets, using a number of real datasets and production level algorithms.

## 1.2. Evaluation of Probability of Default models

The evaluation of PD models is a well studied topic. The quantitative part of the evaluation can be broken into 3 different stages: calibration, discrimination and stability (Castermans, 2009). Discrimination and calibration are measures that determine how well the estimated PDs fit the data, but while discrimination measures how well the rating system provides an ordinal ranking of the risk measure considered, calibration measures the quality of the mapping between a rating and the PDs. Stability measures to what extent the population that was used to construct the rating system is similar to the population on which it is currently being used (Madema, 2009, Castermans, 2009).

Common metrics for calculating calibration are the Brier Score Loss, the Binomial test, the Chi-squared test, the Traffic Lights Approach, and the Hosmer-Lemeshow test (Allison, 2014; Glennon, 2008; Medema, 2009; Committee on Banking Supervision, 2005; Engelmann, 2011). For the context of this paper we will only use the Brier Score Loss, which allows us to evaluate classifiers without the influence of the creation of risk classes (grade “pools”).

The act of calibrating can be defined as learning a function that maps the original probability estimates, or scores, into more accurate probability estimates (Bostrom, 2008). A classifier is considered well calibrated if the set of individuals to which it attributes a probability P of belonging to the positive class (in our case of defaulting on a loan) are indeed (on average) P% likely to belong to that class (Kleinberg, 2016).

### 1.2.1. Brier Score

The Brier Score (or Brier Score Loss) is a measure of calibration, defined as

$$B = \frac{1}{N} \sum_{i=1}^{N} (\hat{p}_i - Y_i)^2$$

Where $\hat{p}_i$ is the estimated probability of the observation $i$, and $Y_i$ is its observed (actual value) (Medema, 2009). Should we wish to use pooled loans (where risk classes have been attributed) we can use the following alternative:

$$B = \frac{1}{N} \sum_{k=1}^{K} N_k [p_k (1 - PD_k)^2 + (1 - p_k)(PD)_k^2]$$

Where $PD_k$ denotes the probability of default assigned to each obligor in rating grade $k$ and $p_k$ is the observed default rate within the same rating grade (Engelmann, 2011).

In a perfectly calibrated model $B$ will be zero, so the metric is sometimes called Brier Score Loss.

### 1.2.2. Area under the Receiver Operator Characteristic curve (ROC Curve and AUROC)

The Receiver Operator Characteristic Curve (ROC curve) is highly used to evaluate binary classifiers in machine learning applications (Foster, 2003). It is a measure of rank, estimating the probability that a random positive is ranked before a random negative, without committing to a particular decision threshold (Flach, 2011).

The ROC Curve is built by plotting the proportion of points correctly classified as class 0 (True Negative Rate) on the vertical axis against the proportion of points incorrectly classified as class 1 (Adams, 1999). In credit risk terms, we have:

> **Vertical axis:** proportion of loans that were correctly classified as default (*"as predicted, they defaulted"*).

**Horizontal axis:** proportion of loans that were classified as defaulting, but ended up being safe (*"they were safe, despite being predicted as defaults"*).

Naturally, an ideal ROC curve would go perfectly vertical until having correctly classified all of the class 1 point (defaults) as being class 1.

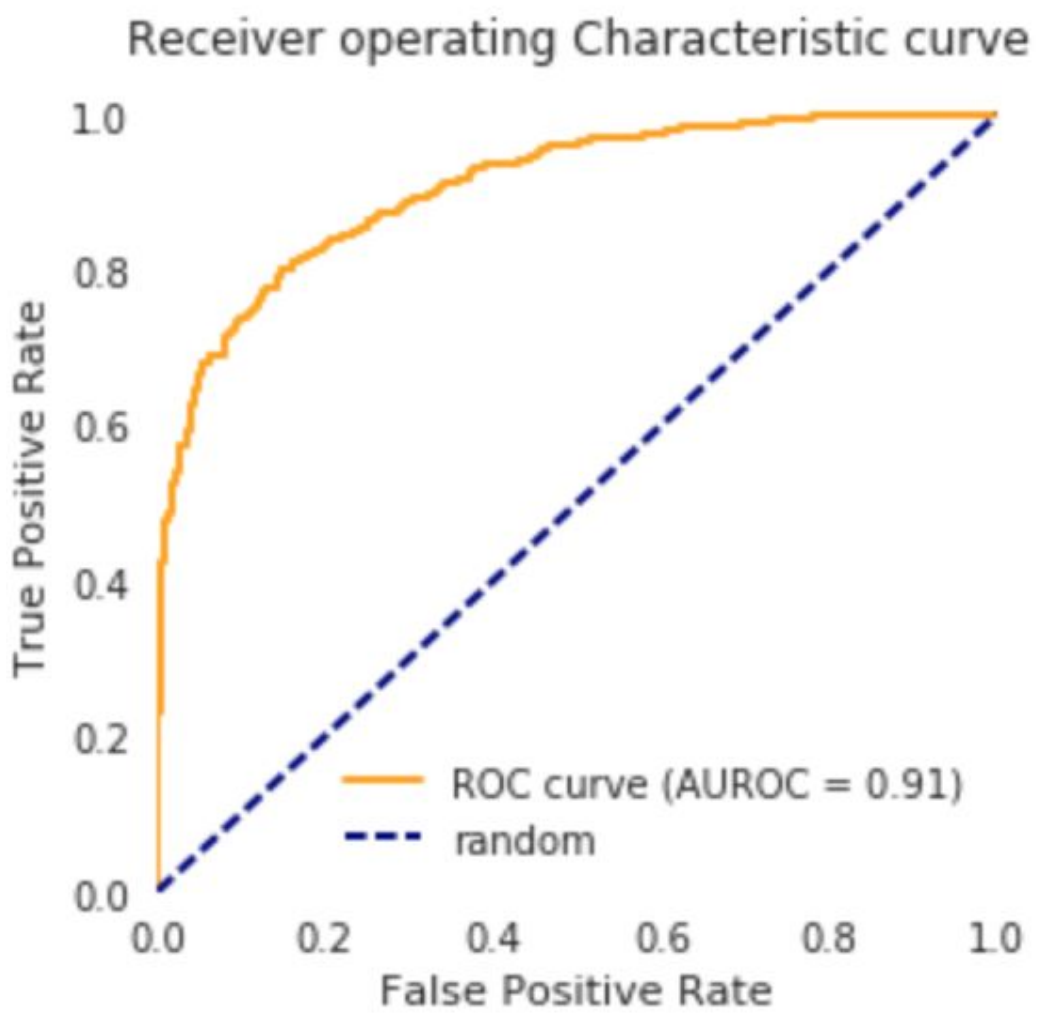

*ROC curve*

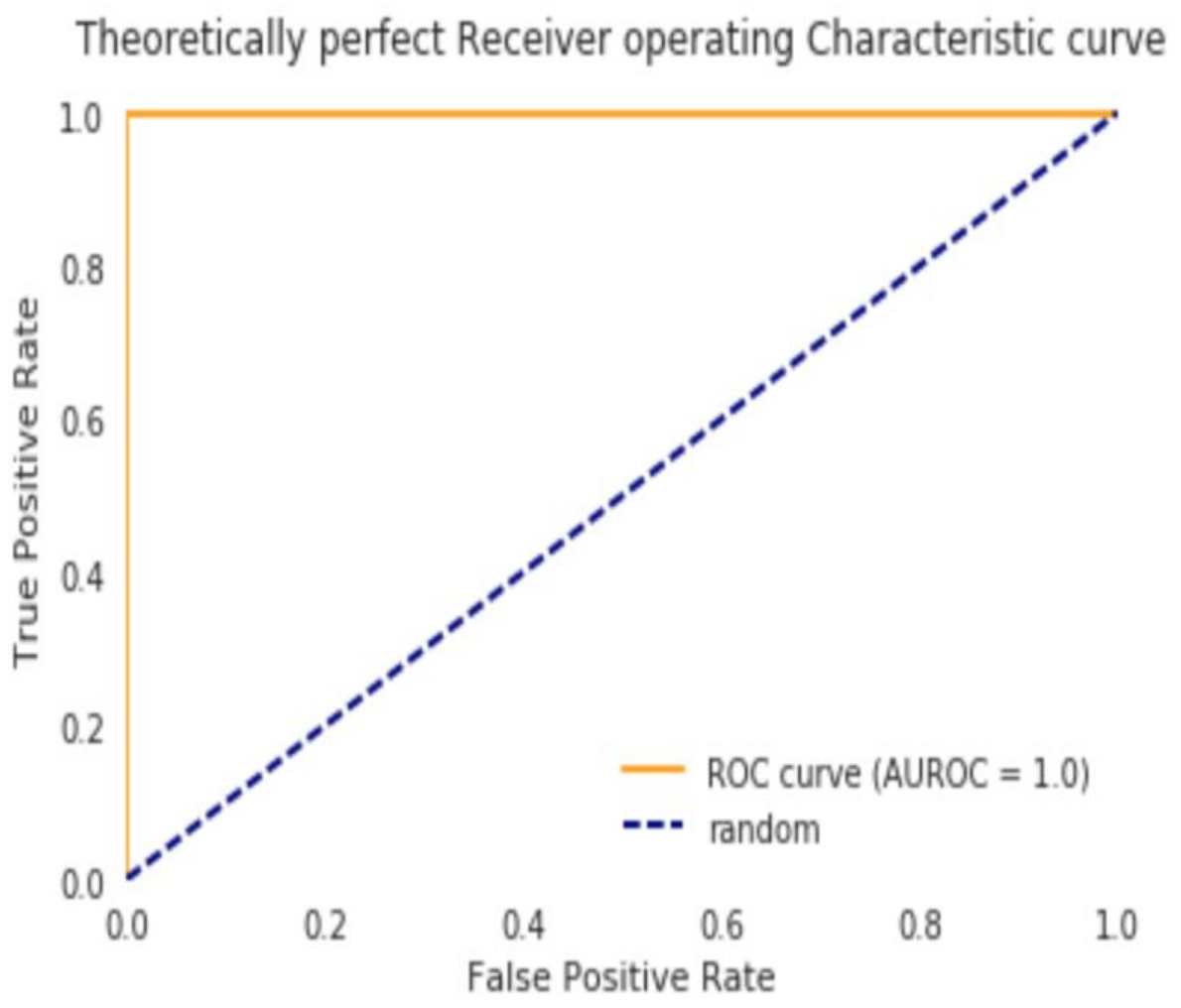

Theoretically *Perfect ROC curve*

In order to get a measure of rank with a single number, the Area Under the ROC curve (AUROC) is often used (Bradley, 1997). This can be calculated with trapezoidal integration (Bradley, 1997), and will be equal to 1 in a perfect model. If the model is no better than random, then the AUROC will be 0.5. A negative AUROC indicates that the model can have predictive power if the decisions are reversed.

## 1.2.3. Gini Coefficient

The Gini Coefficient is frequently used in credit scoring as a replacement for AUROC, and is simply the linear transformation of the AUROC, standardised so that the chance (random) classification has a score of 0 (Hand, 2013; Adams, 1999). The conversion formula is given by

$$Gini = 2\ AUROC - 1$$

The Gini Coefficient can be seen as the quotient of the area which the Cumulative Accuracy Profile curve and diagonal enclose, and the corresponding area in an ideal rating procedure (Engelmann, 2003). The conversion between AUROC and Gini Coefficient, according to the formula above is depicted in the following figure.

*Gini vs Auroc Table*

| AUROC | Gini |
|---|---|
| **0.0** | -1.0 |
| **0.1** | -0.8 |
| **0.2** | -0.6 |
| **0.3** | -0.4 |
| **0.4** | -0.2 |
| **0.5** | 0.0 |
| **0.6** | 0.2 |
| **0.7** | 0.4 |
| **0.8** | 0.6 |
| **0.9** | 0.8 |
| **1.0** | 1.0 |

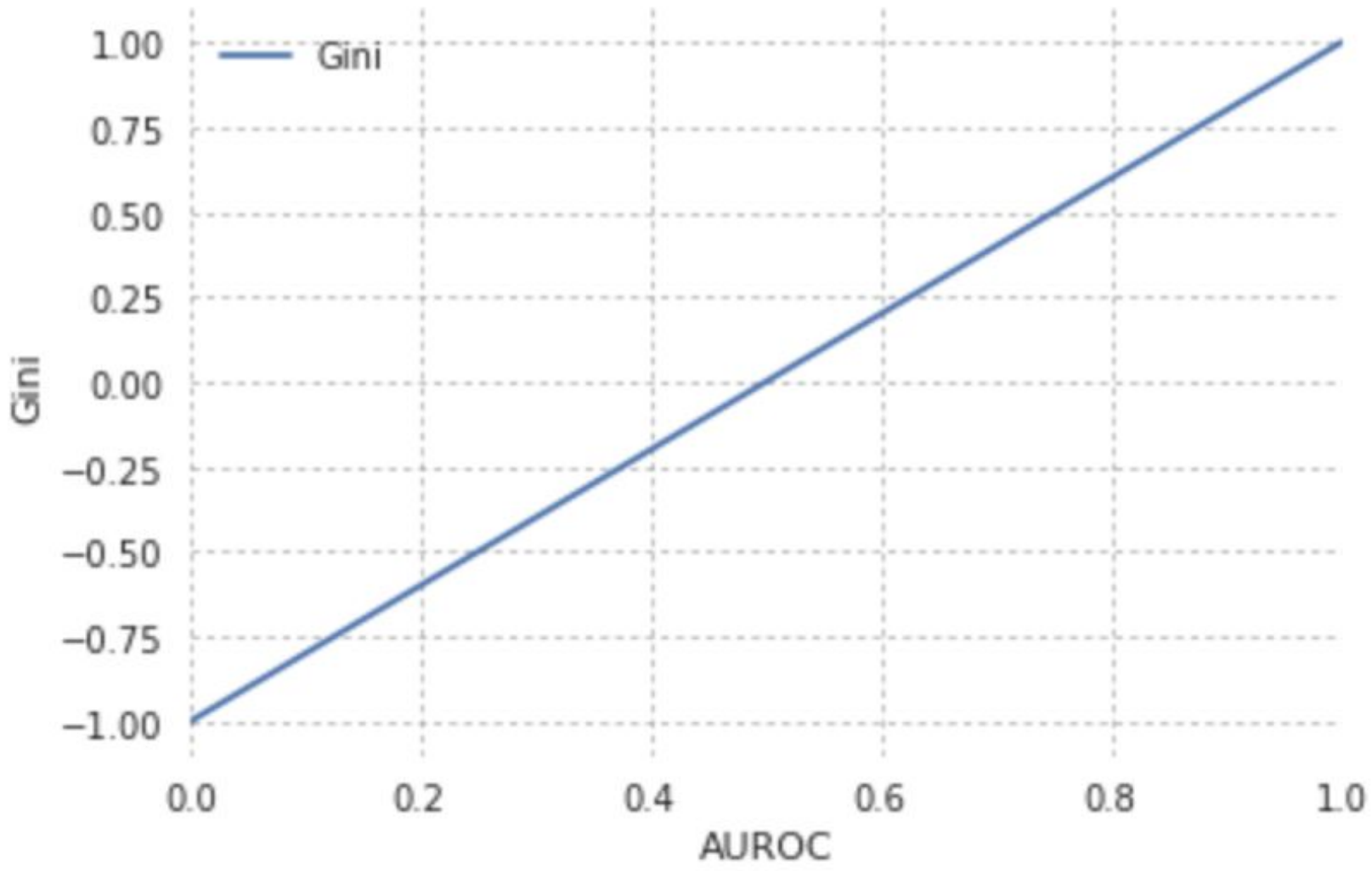

*Gini vs AUROC*

Alternatively, a Corrected Gini Coefficient is also common and it is the summary statistic of the Lorenz curve: the ratio of the area between the perfect equality diagonal and the blue line (Lorenz curve) and the area below the perfect equality diagonal (Izzi, 2012). Nevertheless, this paper focuses on the former definition of the Gini.

### 1.2.4. Limitations of evaluating a classifier using only AUROC and Gini

AUROC and Gini are very useful metrics for evaluating the rank ability of a model. However, relying only on rank order has serious limitations, because it ignores the ability of the classifier to accurately identify probabilities. As an example, consider a toy dataset (see Annex 1) where we have some predictions of probabilities (between 0 and 1) and some observed results for those predictions (either 0 or 1).

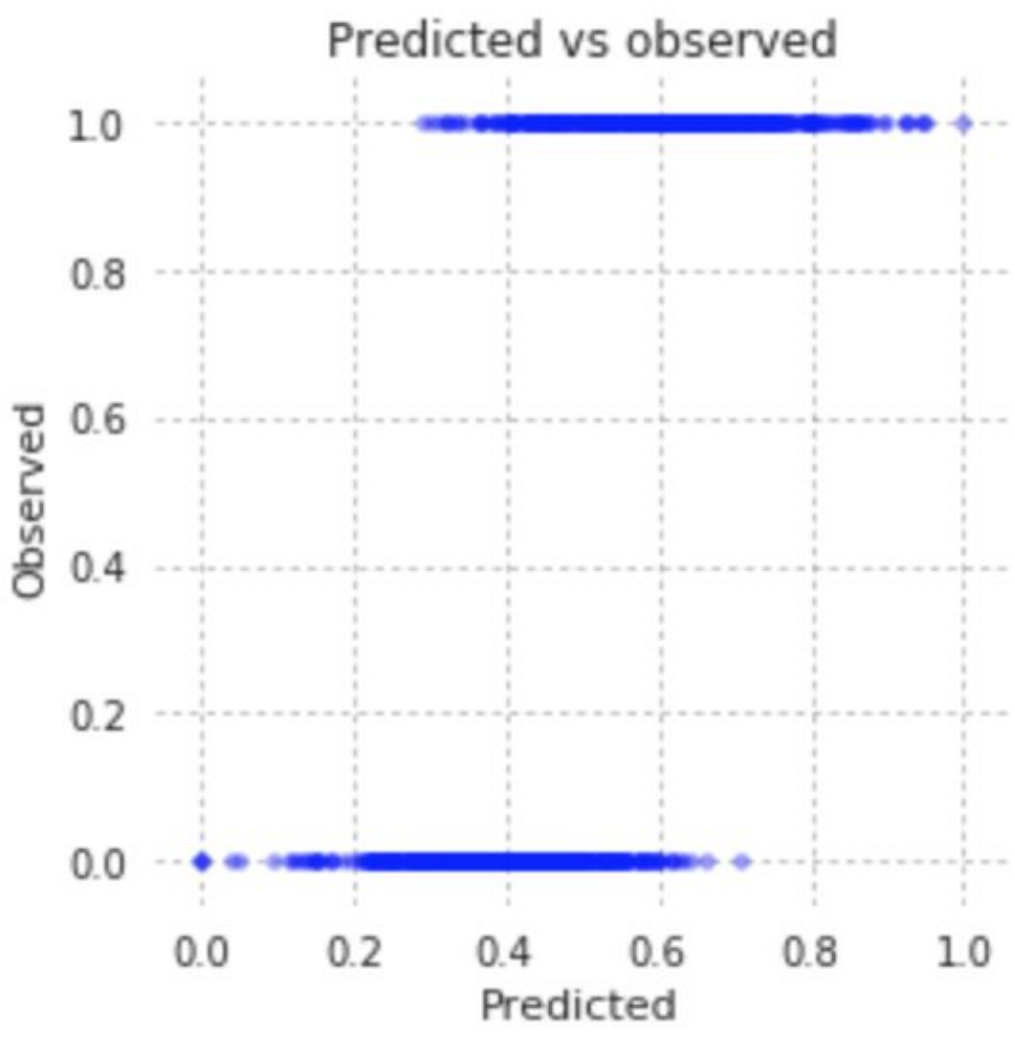

*Predicted (PD) vs observed (Default rate)*

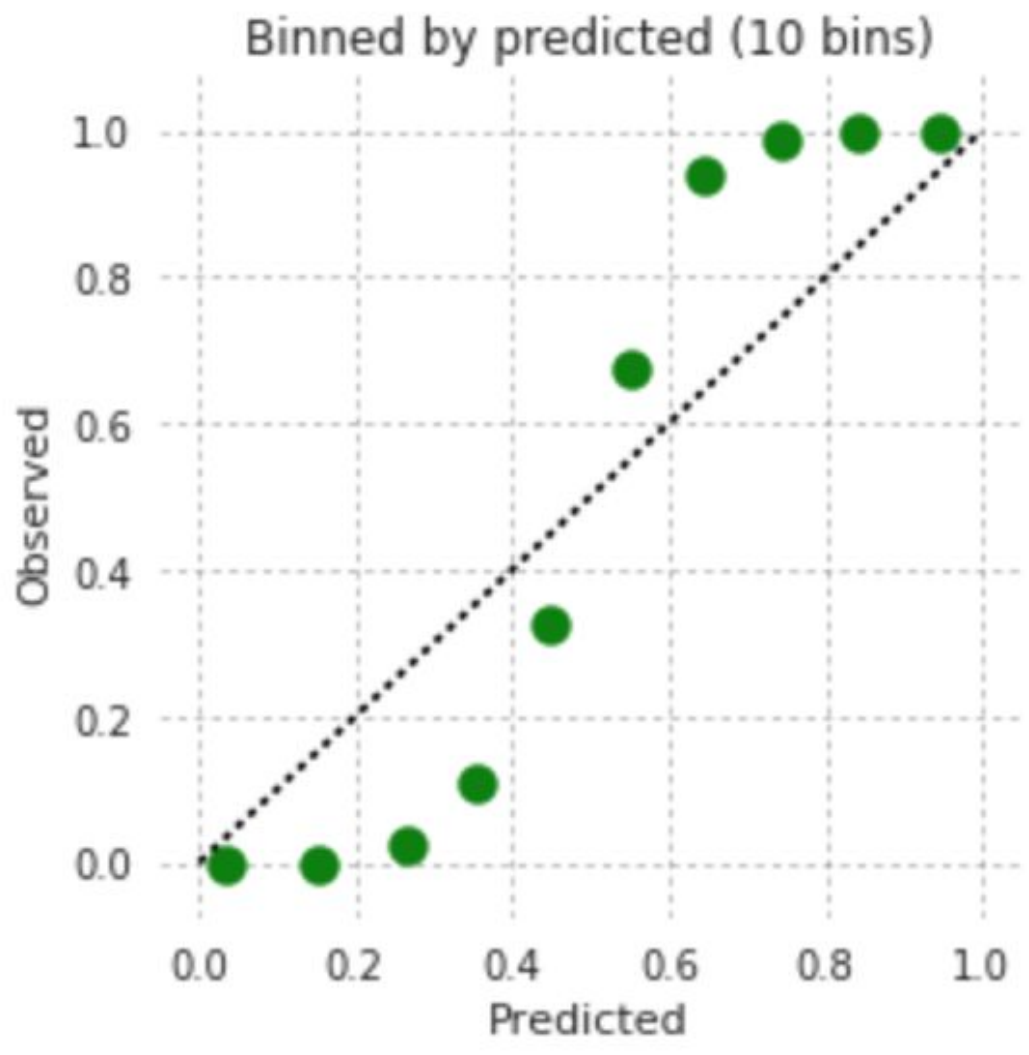

*Predicted (PD) vs observed (Default rate), with 10 bins*

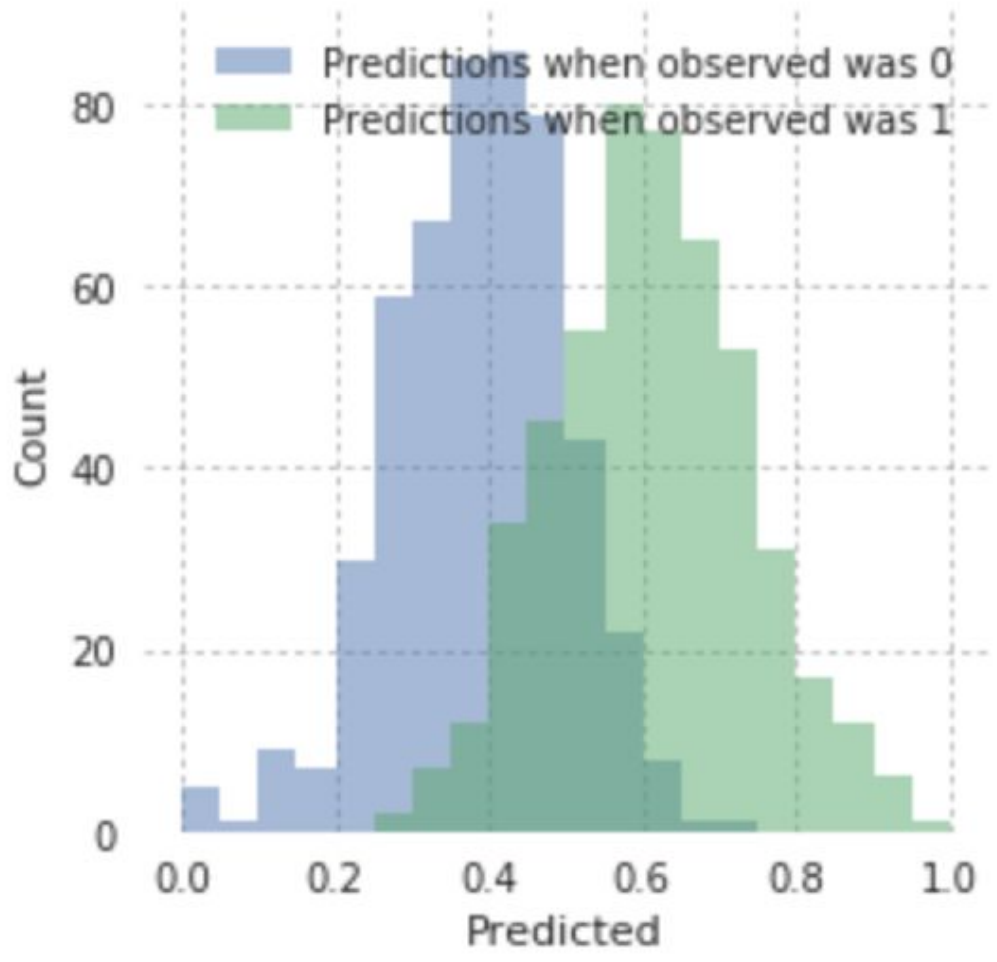

*Predictions (PD) histogram*

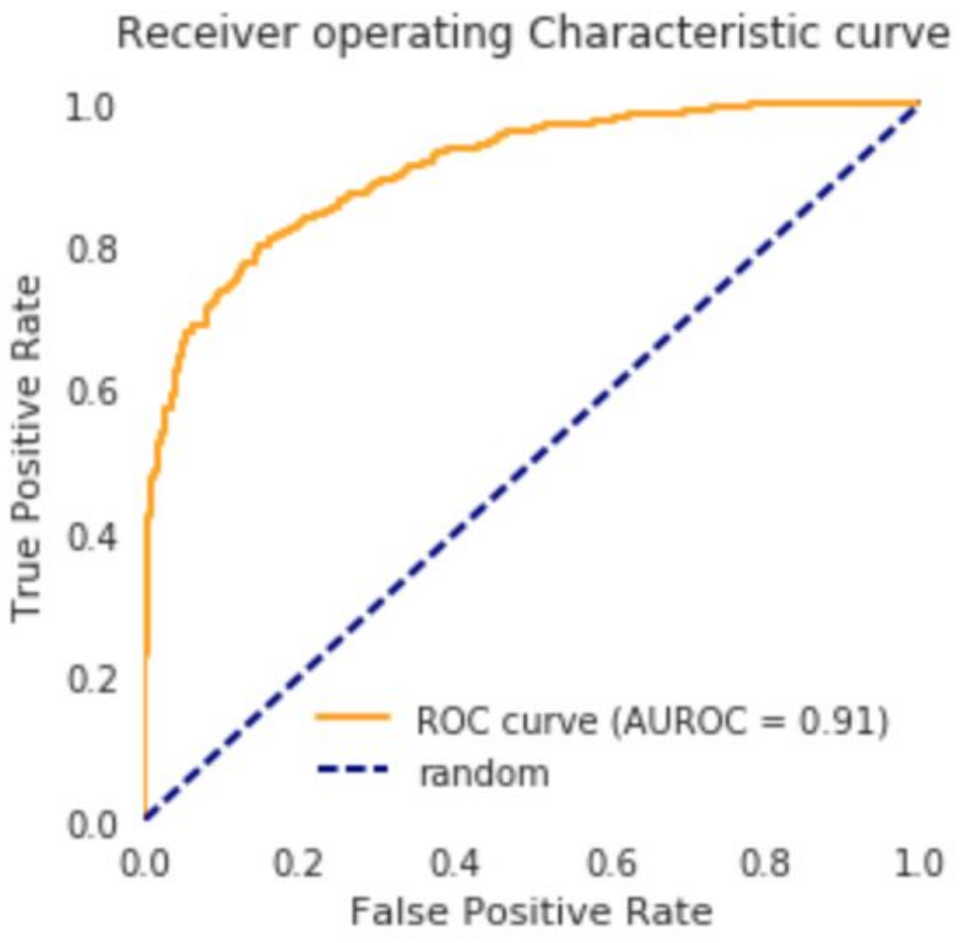

*ROC Curve*

For this dataset, we have the following:

```
Roc Auc:            0.91
Gini:               0.81
Brier score loss:   0.17
```

Now, if we take all of our predictions and divide them by two (clearly making our accuracy worse), we will get the following:

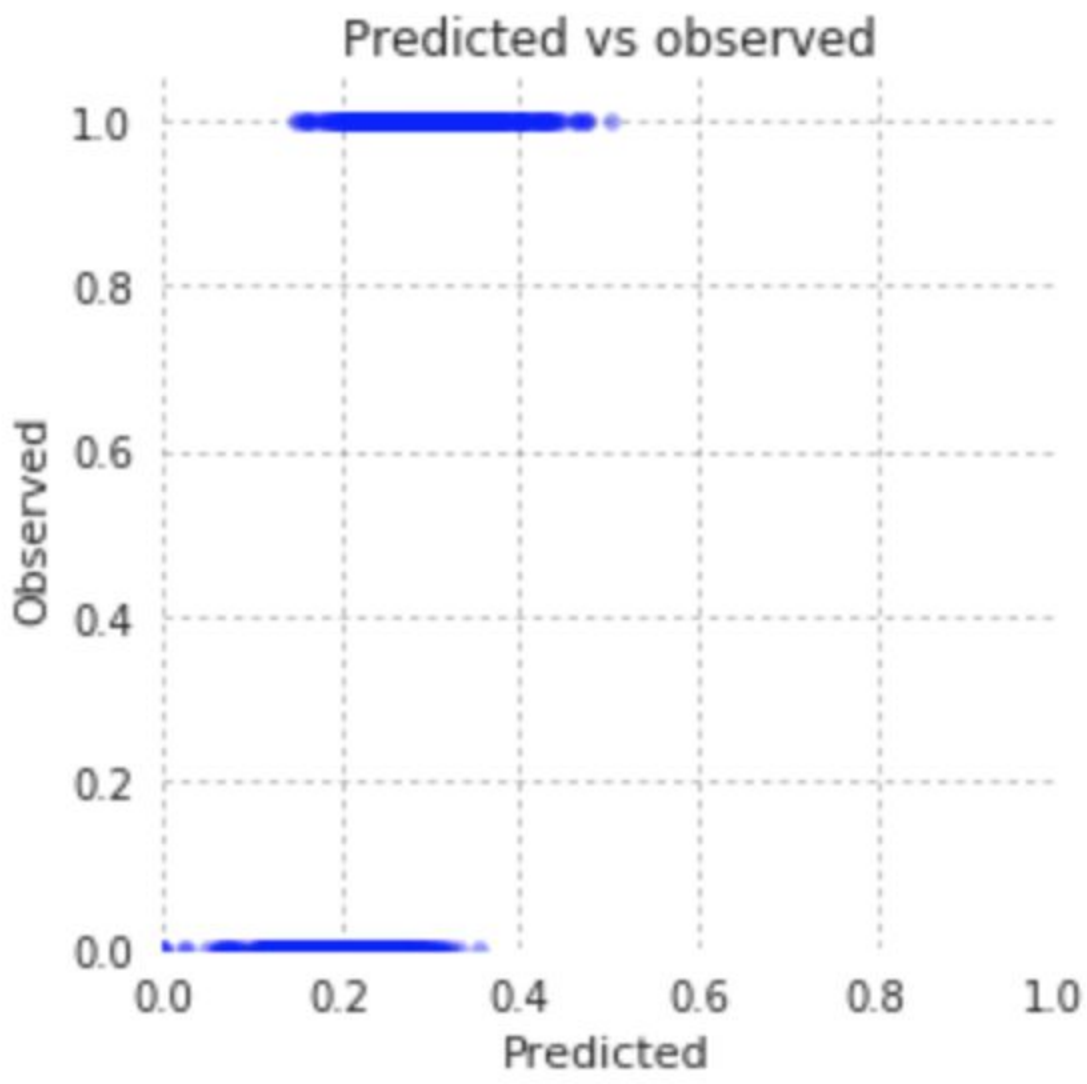

*Predicted (PD) vs observed (Default rate)*

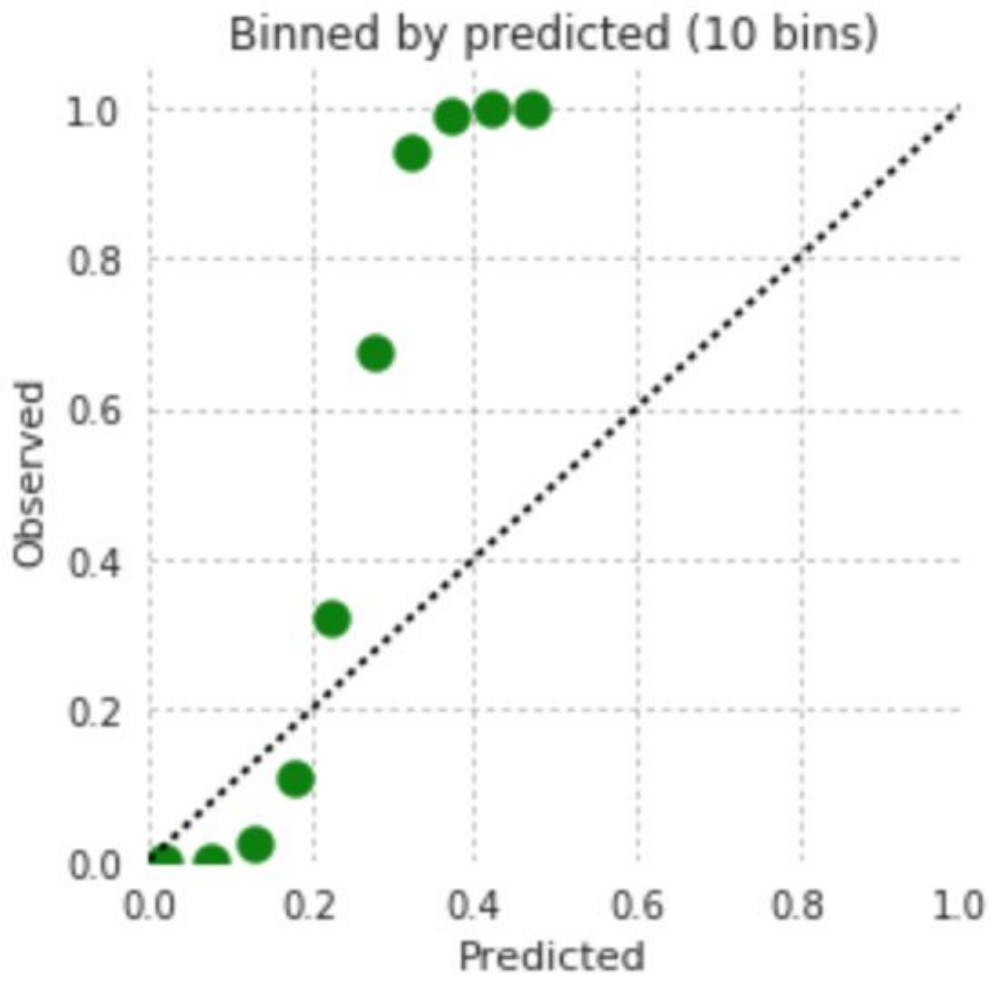

*Predicted (PD) vs observed (Default rate), with 10 bins*

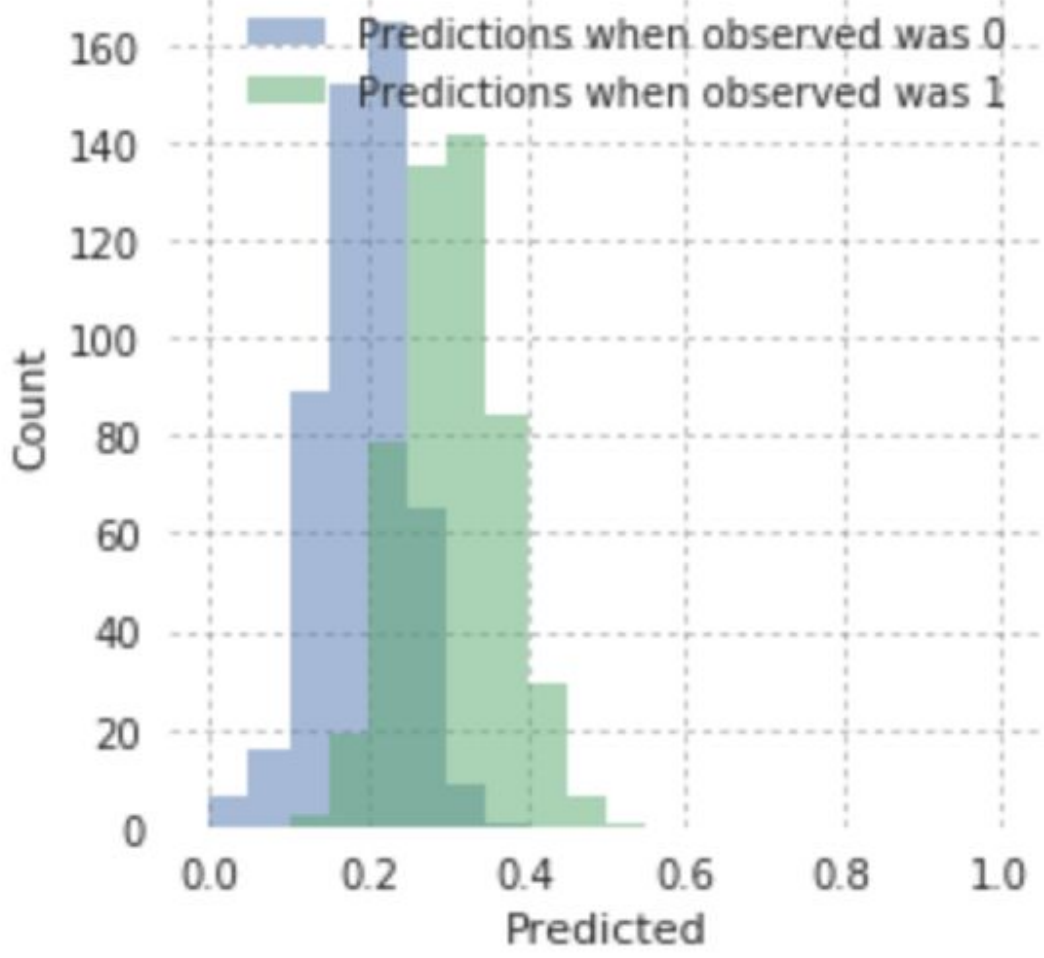

*Predictions (PD) histogram*

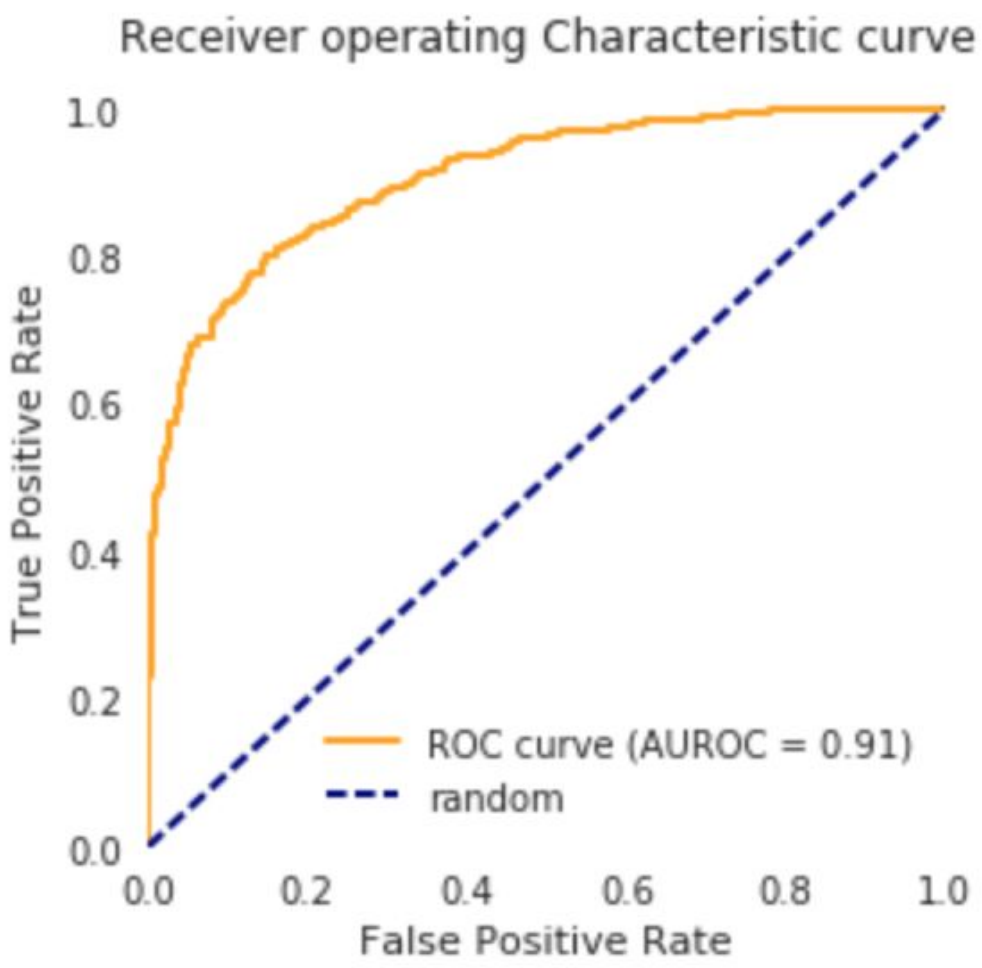

*ROC Curve*

```
Roc Auc:          0.91
Gini:             0.81
Brier score loss: 0.26
```

The ROC Curve and AUROC are in no way affected, because we preserved the rank ability of the predictions. However the Brier score loss has captured this difference, and has changed from 0.17 to 0.26.

## 1.3. Machine Learning Algorithms in Credit Risk

A number of studies have shown the advantages of using machine learning systems in credit scoring problems, and how they can achieve superior performance to the traditional scorecard-based approaches (Nanni, 2009; Lessmann, 2015). The emergence of these methods in open source libraries (such as Scikit-learn, R or Weka) and in proprietary software solutions (e.g. SAS) has made them widely available to the general population and to the lenders themselves.

Benchmarks of machine learning classifiers on credit scoring datasets have been comprehensively cataloged by (Lessmann, 2015), but these benchmarks have generally focused on rank ability, generally measuring the Gini score or AUROC, rather than on calibration.

## 1.4. Calibration of machine learning algorithms

Calibration of machine learning algorithms can be done using various approaches. The most common are Platt Scaling and Isotonic Regression (Platt, 2000;  Boström, 2008), which we shall be exploring in this paper.

### 1.4.1. Platt scaling

Platt scaling, often called sigmoid scaling, is most effective when the distortion in the predicted probabilities is sigmoid shaped (Niculescu-Mizil, 2006).

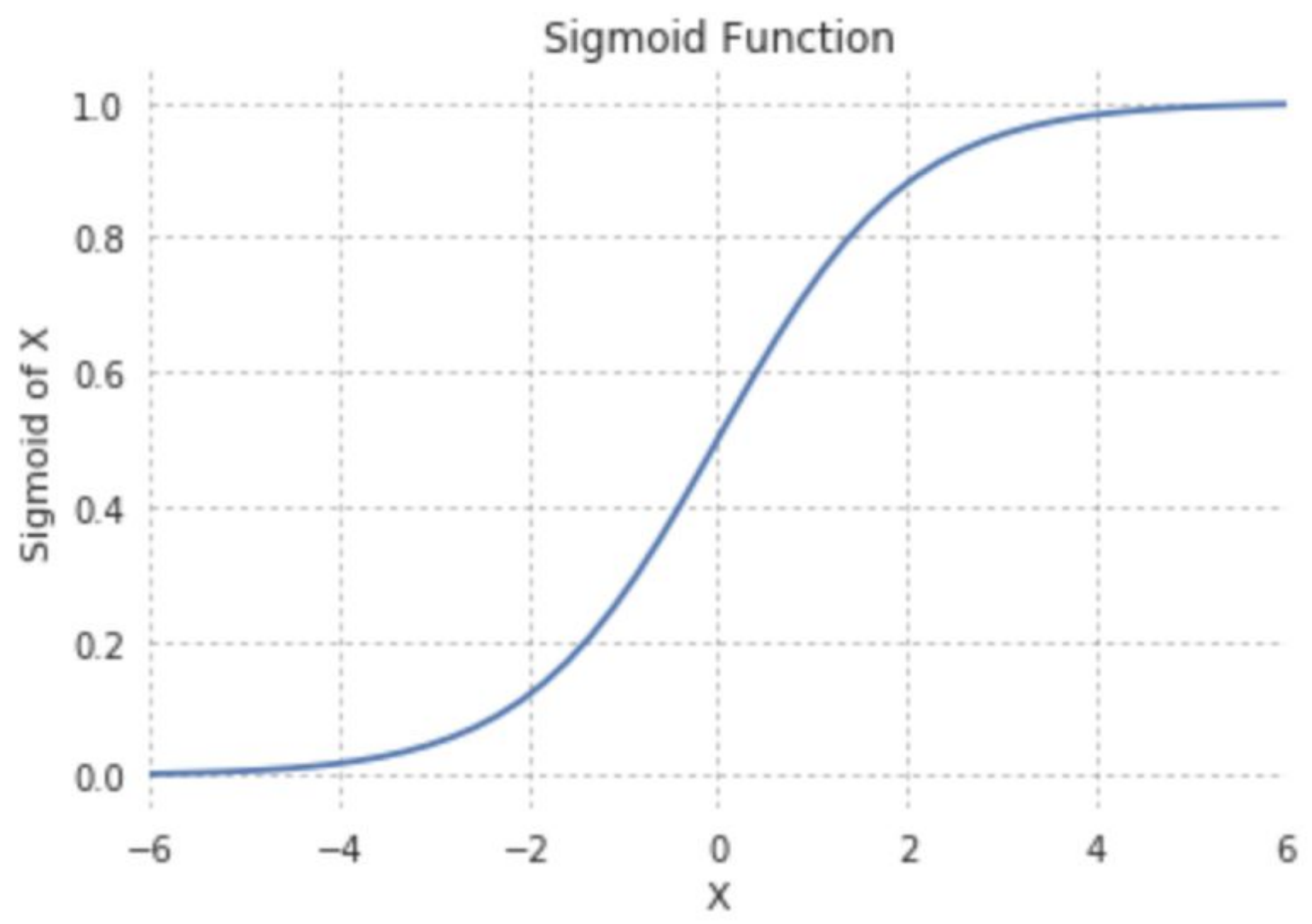

*Example sigmoid function*

The output of the classification system (PDs or scores) are passed through a sigmoid, where the parameters A and B are fitted using maximum likelihood estimation from a fitting training set (Platt, 2000), given by

$$P(y = 1|f) = \frac{1}{1 + exp(Af + B)}$$

This fitting is done by using gradient descent (Niculescu-Mizil, 2005) to minimize

$$argmin_{A,B}\{-\sum_{i} y_i log(p_i) + (1 - y_i) log(1 - p_i)\} \text{ , where } p_i = \frac{1}{1 + exp(Af + B)}$$

To avoid overfitting, the parameters should be fit on an set which is different from the one where the model was initially fit. More advanced approaches include the use of cross-validation for using the training set to both fit and calibrate the classifier (Hawkins, 2003), which is beyond the scope of this paper.

### 1.4.2. Isotonic regression

As defined in (Zadrozny, 2012) isotonic regression is *a non-parametric form of regression in which we assume that the function is chosen from the class of all isotonic (i.e. non-decreasing) functions.* It can be considered a general form of binning, with the advantage that it does not require any specific number of bins to be predetermined or any limits of the size of each bin (Boström, 2008).

Given predictions $f_i$ from our classifier, and the true target $y_i$, the isotonic regression is simply

$$y_i = m(f_i) + \varepsilon_i \text{ ,}$$

where $m$ is a non-increasing function (Niculescu-Mizil, 2015). This allows for greater flexibility, but at the cost of being more prone to overfitting (Menon, 2012). In general, Isotonic regression will outperform Sigmoid regression only when there is enough data to avoid overfitting (Niculescu-Mizil, 2005).

# 2. Methodology

## 2.1. Benchmark

### 2.1.1. Overview

The benchmark aims to test the effect of re-calibrating on consumer credit datasets. The calibration was evaluated using Brier Score Loss. For each dataset, three different machine learning models were fit, and the model was calibrated using sigmoid and isotonic regression. One extra control run was made without any calibration. The benchmark was run on a total of 18 datasets from different banks and non-bank lenders, who allowed the use of their meta-data for the production of this study, together with publicly available consumer credit datasets from online lending platforms. The datasets are of consumer loans, and the target is whether the loan defaulted.

For each experiment, the data was partitioned into 3 sets. The model was trained on the chronologically first 60% (training set), and calibrated on the next 20% (calibration set). Another set (20%) was left separate to act as a “recent” dataset, after the calibration.

| Training set | Calibration set | Recent data set |
|---|---|---|
| 60% | 20% | 20% |

This design of experiments led to the following experiments, for each of the 18 datasets, leading to a total of 162 experiments, with the following nomenclature:

| | Logistic Regression | Random Forest | Gradient Boosting Classifier |
|---|---|---|---|
| None | E1 | E2 | E3 |
| Sigmoid | E4 | E5 | E6 |
| Isotonic | E7 | E8 | E9 |

For each dataset, for each for these experiments, the following steps were run:

1. `Partition dataset into Training, Calibration and Recent data sets`
2. `Train on Training set`
3. `Predict probabilities on Training set`
4. `Predict probabilities on Calibration set`
5. `Predict probabilities on Recent data set`
6. `Fit calibration with Calibration set (if using either Isotonic or Sigmoid)`
7. `Predict probabilities on Calibration set`
8. `Predict probabilities on Recent data set`
9. `Measure calibration on all sets`

## 2.2. Type of models used

### 2.2.1. Logistic regression

Unlike some other machine learning models, Logistic regression models' results are already outputted as probabilities (Nosslinger, 2004). Studies suggest that attempting to calibrate Logistic regressions do not improve results (Niculescu-Mizil, 2006), and can be counter productive and move the probability mass away from 0 and 1 when the sigmoid regression is used (Niculescu-Mizil, 2005) .

### 2.2.2. Random Forest

For a random forest of classification trees, the probability distribution is formed by averaging the unweighted class votes by the members of the forest, where each member vote for a single (the most probable) class.

Random forests are ensemble models that use classification trees, where the probability distribution is formed by averaging the unweighted class votes of each member tree (Bostrom, 2008). Each of the trees is only allowed to use a fraction of the training data and a fraction of the features (Ben-David, 2009), making the Random Forests more resistant to overfitting than classical Decision Tree Classification Models. Literature is less coherent on the calibration of random forests than on the calibration logistic regression, with (Bostrom, 2008) claiming that they are relatively well calibrated, (Niculescu-Mizil, 2006) claiming that they produce good probabilities, whereas studies by (Li, 2013) and (Dankowski, 2016) showing that there is room for improvement.

### 2.2.3. Gradient Boosting Classifier

The gradient boosting classifier is an additive model, where regression trees are fit on the negative gradient of the loss function (Friedman, 2001). (Niculescu-Mizil, 2006) considered that calibration with either Platt scaling or Isotonic scaling could improve their performance, and that after calibrated, boosted trees yielded excellent results.

## 2.3. Data

The 18 datasets used were consumer lending datasets, whose results were aggregated to preserve anonymity.  The type of datasets vary considerably. The length of the time series (shown here in weeks) goes from a few weeks up to a few years.

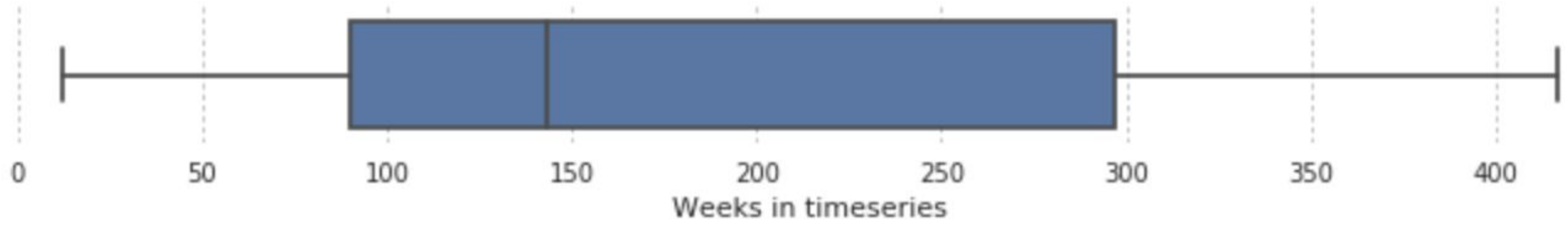

The default rate was generally under the 20% mark, with the median around 6%.

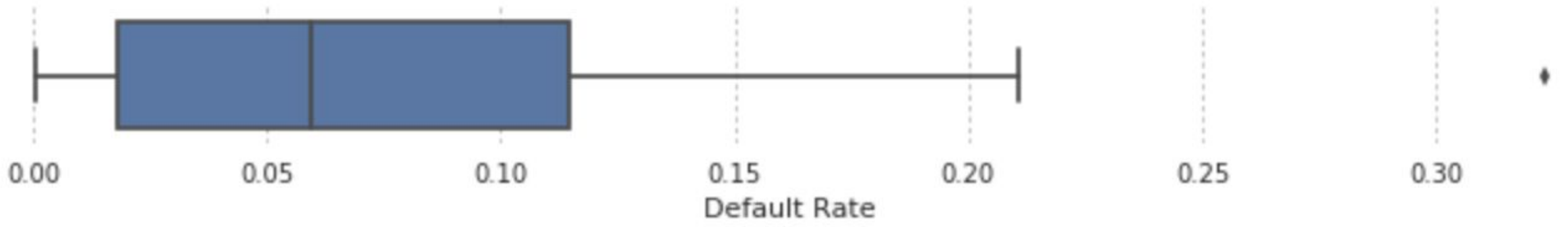

The number of features was generally under 100, with a few lenders choosing very high numbers of features.

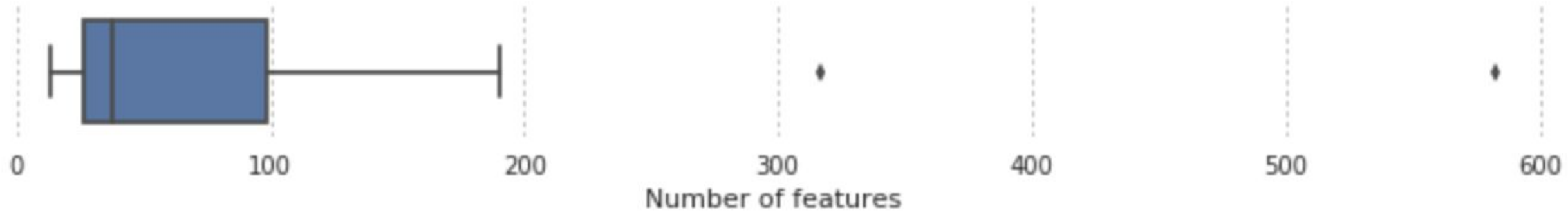

The amount of data was typically on the order of tens of thousands per dataset, with a few exceptions in the hundreds of thousands.

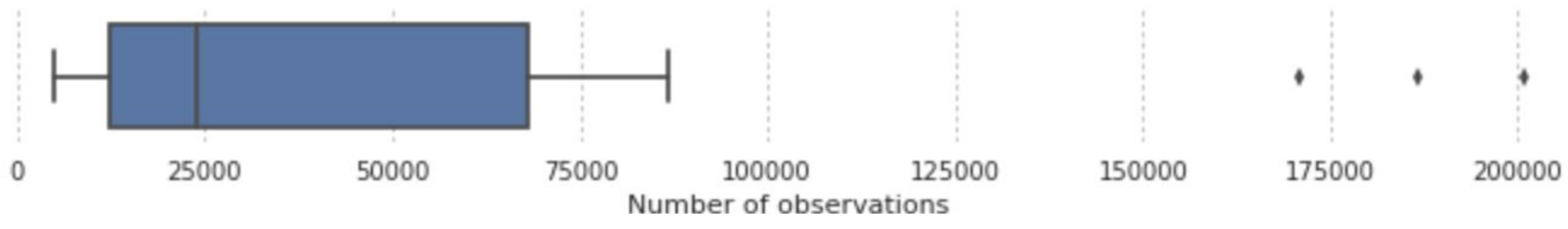

# 3. Results

Let's start by observing the brier scores on the training set of each dataset. The following chart shows the boxplot of Brier scores for each classifier, without any re-calibration, on the training set:

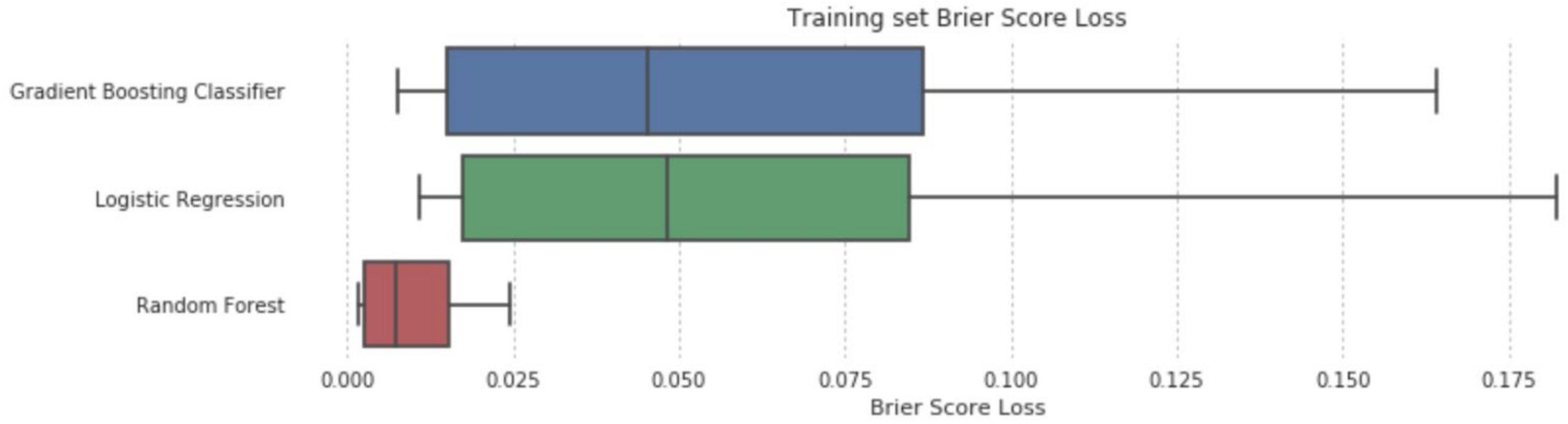

The Random Forest has a very low brier score on the training set, but this is naturally not very meaningful. Let's observe the brier score on the test set (last 20% of each dataset), for each classifier, without any re-calibration:

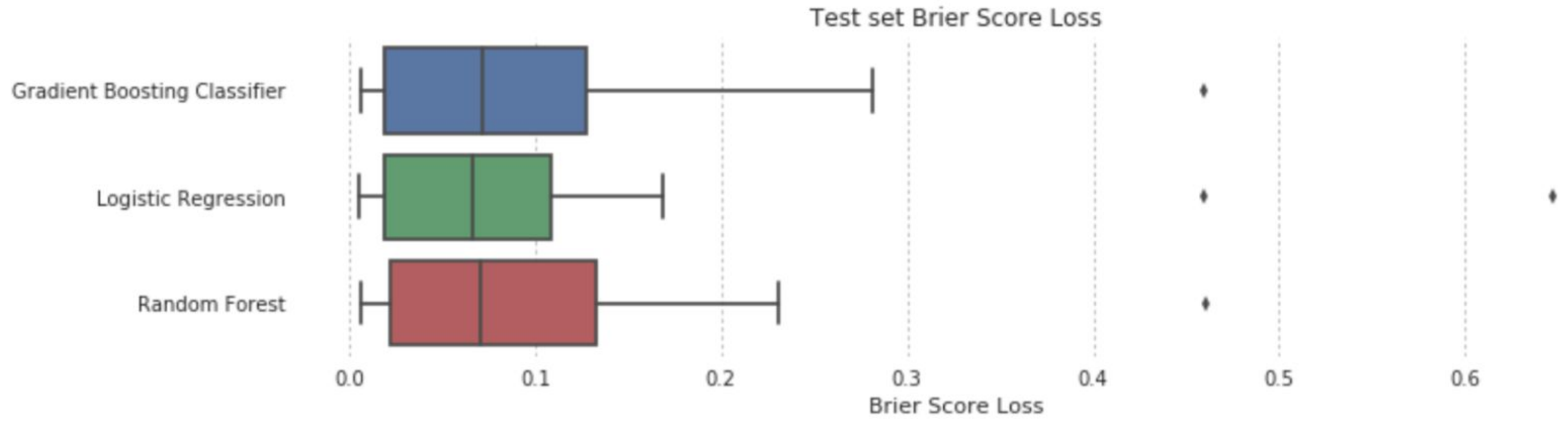

Clearly all classifiers have lost calibration ability. We can control for the natural variation of the datasets by doing the following:

```
For each dataset:
        Get the brier score for:
```

```
                    The Logistic Regression (brier-logit)
                    The Random Forest Classifier (brier-rf)
                    The Gradient Boosting Classifier (brier-gbc)
             (brier-rf-norm) = (brier-rf) / (brier-logit)
             (brier-gbc-norm) = (brier-gbc) / (brier-logit)
        plot (brier-rf-norm) and (brier-gbc-norm)
```

Having controlled for the dataset (via the score on the logit), we can now see how each classifier behaved on the test set.

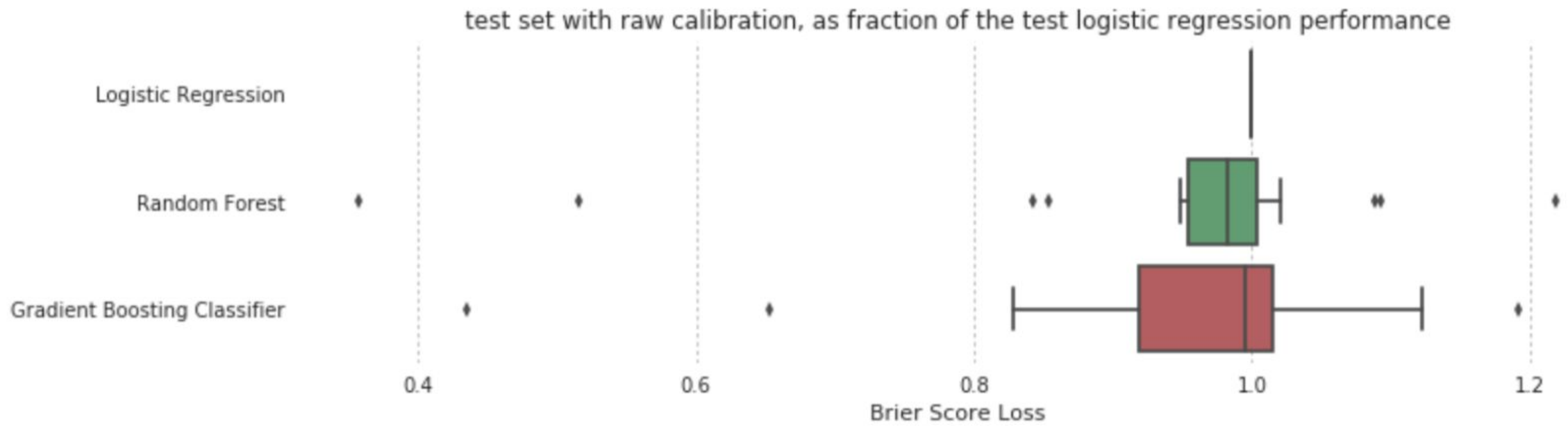

The results are a lot more informative, and actually contradict our initial impression from the previous chart: on most datasets, the Random Forest and Gradient Boosting actually outperforms the Logistic Regression on Brier Scores.

With sigmoid calibration, we get some improvements. As a reminder, the model is trained on the first 60% of the data, the calibration was done on the data between percentile .6 and .8, and the test set is the last 20%, along a time series.

The results are divided by the performance of the classifier on the particular dataset, to control for natural variation:

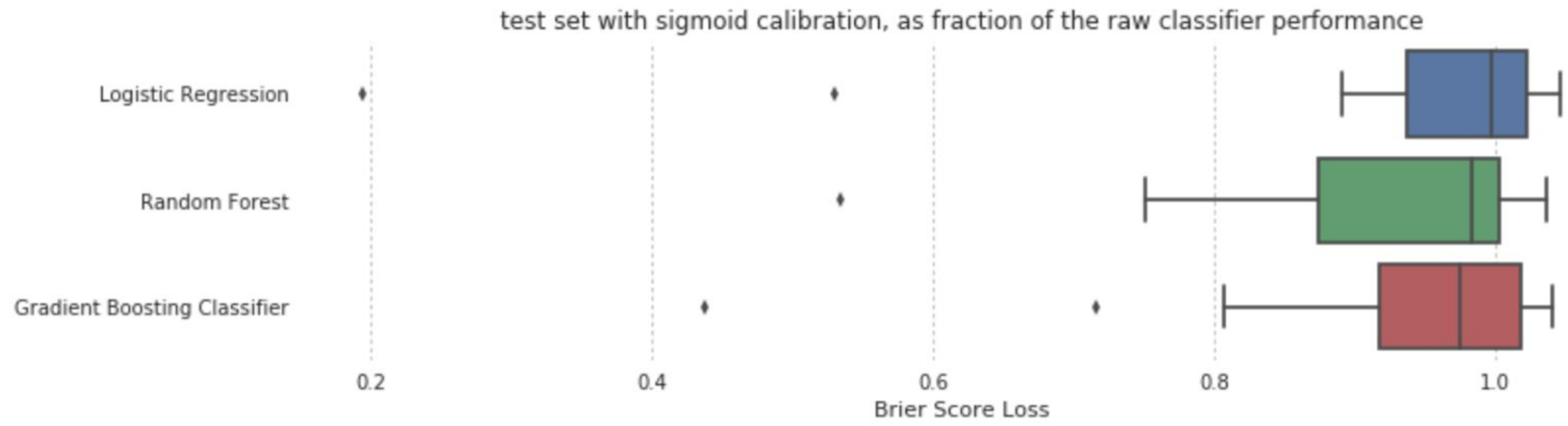

We can also see them as a percentage of the raw logistic regression performance on the test set:

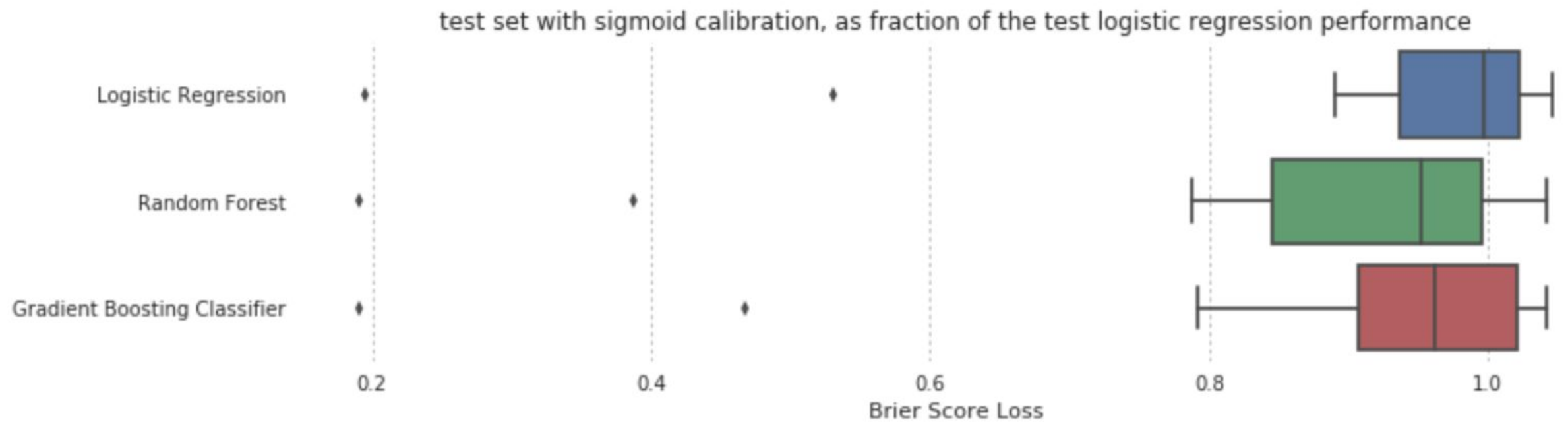

It would appear that the Random Forest is the one that has the most to gain from calibration with Sigmoid. However it should be noticed that in about 25% of cases calibration it did more harm than good. Logistic regression does not seem to benefit at all from sigmoid calibration, and Gradient Boosting seems to be improved in most cases, but slightly damaged in slightly over 25% of cases.

How about Isotonic Calibration? Once again, we start by comparing the performance with the uncalibrated performance, on the test set:

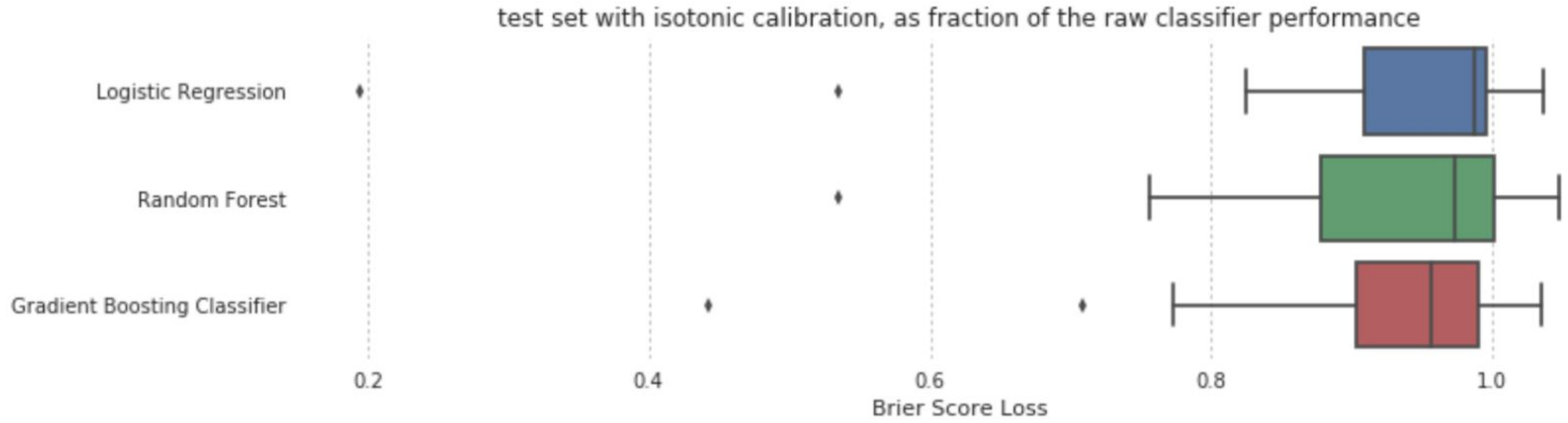

And add the performance as a fraction of the raw logistic regression performance, for reference:

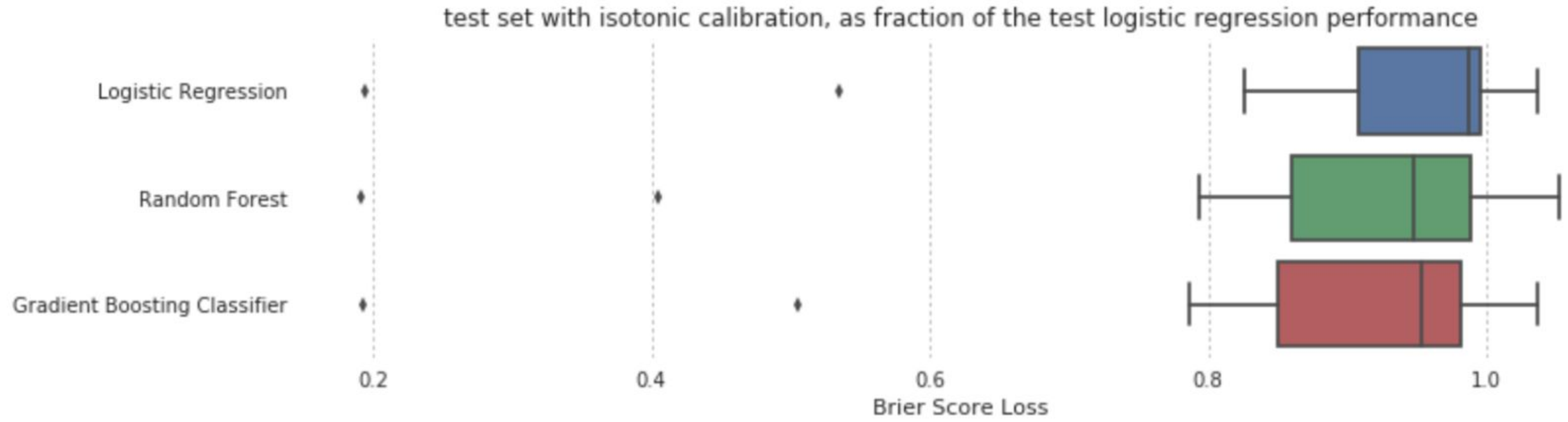

The first thing we notice is that, unlike with sigmoid calibration, isotonic calibration can actually help the logistic regression. It also slightly outperforms the performance of the calibrated Random Forests, and helps the gradient boosting classifier. Interestingly, both the Random Forest and Gradient Boosting Classifier clearly outperform the logistic regression after calibration.

# 4. Discussion

The objective of this paper was to provide a benchmark of calibration with different datasets, different classifiers, and different calibration techniques using real world consumer credit datasets. The amount of datasets any study has access to is evidently limited, but the results seem sufficient to arrive at a few tentative conclusions.

Firstly, that calibrated non-parametric models seem to on average provide better calibration across a time series than the simple or calibrated logistic regression. On the other hand the non-calibrated non-parametric models do not perform much better than the logistic regression, and can actually yield considerably worse performance.

Secondly, that the non-parametric models were frequently improved (on average) by re-calibration, but that re-calibration can in a few cases decrease the performance on the test set. This might be explained by observing how different the calibration set was from the test set, or by understanding whether the calibration set had enough data, which would be an interesting scope for further research. Random Forest seems to require more calibration than the Gradient Boosting classifier.

Thirdly, that the re-calibration of all models is generally best achieved with isotonic calibration, rather than with sigmoid (Platt) calibration. This is particularly true of the Logistic Regression, where the Platt calibration yielded almost no improvement on average, while the Isotonic calibration improved more than 75% of cases, something by quite significant amounts.

Fourth, that testing calibration on the training set can be very misleading, particularly as non-parametric models seem to be excellently calibrated on the test set, while not performing similarly on the test sets. As with other metrics in machine learning, calibration should be evaluated on a disjunct set from the one on which the model was trained.

# 5. Further work

A number of topics are interesting for further investigation. It would be useful to understand the impact of the size of the calibration set on the ability of the isotonic regression to not overfit, and to generalize to a separate test set. Learning curves would be very helpful with understanding this phenomenon. Larger experimental grids should be attempted, as there is no guarantee that we chose the right set of hyper parameters to improve calibration, and it reasonable to assume that a careful choice of hyper parameters can improve calibration considerably.

Another interesting field of investigation would be to see how the similarity between the calibration and test set (measured in different ways) correlates with the calibration performance in the test set. Finally, the ability to use cross validation, which is made available in packages such as scikit-learn and R would be worth exploring further, to be able to use the training set both for training and calibrating, and benchmarking against the naive approach used in this paper.

As a last consideration, since most lenders in practice use scorecards, the re-calibration of scorecards with different calibration approach would be a very interesting topic for further research.

# 6. Annex 1: Building the toy dataset

```
import pandas as pd
import numpy as np
def generate_data(corr=0.85, size=1000):
    np.random.seed(5)
    xx = np.array([0, 1])
    yy = np.array([0, 1])
    means = [xx.mean(), yy.mean()]
    stds = [xx.std() / 3, yy.std() / 3]

    covs = [[stds[0]**2          , stds[0]*stds[1]*corr],
            [stds[0]*stds[1]*corr,           stds[1]**2]]

    m = np.random.multivariate_normal(means, covs, size).T

    predicted = pd.Series(m[0])
    predicted.loc[predicted < 0] = 0
    predicted.loc[predicted > 1] = 1

    observed = m[1].round()

    data = pd.DataFrame({'Predicted': predicted,
                         'Observed': observed})
    return data
```